\documentclass[a4paper,12pt]{article}

\usepackage{amssymb}
\usepackage{graphicx}
\usepackage{amsmath}
\usepackage{amsfonts}
\usepackage{booktabs}
\usepackage{txfonts}
\usepackage{cases}
\usepackage{cite}
\usepackage{setspace}

\title{\Large A Robust Signal Classification Scheme for Cognitive Radio}

\begin{document}

\maketitle

\begin{center}
Hanwen Cao and J\"{u}rgen Peissig      \\
Institut f\"{u}r Kommunikationstechnik\,(IKT), Leibniz Universit\"{a}t Hannover \\
Appelstr. 9A, 30167 Hannover, Germany        \\
Email: \{hanwen.cao, peissig\}@ikt.uni-hannover.de
\end{center}

\begin{abstract}
This paper presents a robust signal classification scheme for achieving comprehensive spectrum sensing of multiple coexisting wireless systems. It is built upon a group of feature-based signal detection algorithms enhanced by the proposed dimension cancelation\,(DIC) method for mitigating the noise uncertainty problem. The classification scheme is implemented on our testbed consisting real-world wireless devices. The simulation and experimental performances agree with each other well and shows the effectiveness and robustness of the proposed scheme.
\end{abstract}


\section{Introduction}
The current research on spectrum sensing is mainly focused on detecting signals transmitted by licensed primary users\,(PU), such as TV broadcast, wireless microphone\,(WM), which is motivated by the regulator's stringent protection requirement. Due to the continuing innovations in new wireless technologies and business models, it is expected that more and more heterogeneous systems will share the same spectrum resources in the future. For example, apart from the primary TV broadcast, some other technologies, such as IEEE 802.22, ECMA-392, 3GPP LTE and cognitive PMSE\cite{web:cpmse} are targeted at exploiting the TV band white space\,(TVWS) for providing new services. As a result, the challenge arised is not only to protect the legacy PU, but also to optimize the coordination and coexistence among different secondary devices and networks, especially heterogeneous ones. In this context, in addition to the detection of PU's signal, it is also desired to use spectrum sensing for acquiring knowledge on other coexisting networks or devices. For this purpose, we propose a robust signal classification scheme which can identify the major primary and secondary wireless systems coexisting in TV band, which is based on the combination of a group of feature based signal detection algorithms using properly designed decision rules.

We first review the reported signal classification schemes for spectrum sensing in literatures. In \cite{classi:psd_gi:wang2010}, the authors focus on the classification based on signal's power spectrum density\,(PSD) shape. The cyclic prefix\,(CP) is also taken into account for classifying OFDM signals which have similar PSD. Most of the targeted signals are actually not coexisting in the same frequency band, such as UMTS, DVB-S, HiperLAN and IS-95. In \cite{cycl:ofdm_classi_caf:han2008,cyc:chn_alloc:oener,cycl:classi_ofdm_scld:punchihewa}, classification schemes for OFDM signals are proposed which are based on the analysis of their cyclostationarity. Only the general signal models are discussed in these studies without considering the classification of standard-specific signals.
In \cite{cycl:classi_lte_wimax:al-habashna}, a classifier for LTE and WiMax is proposed based on detailed analysis of the cyclostationary features from both CP and pilot sub-carriers. In \cite{cyc:sutton_thesis:2008}, network identification based on embedded cyclostationary signatures in OFDM signal is proposed. However, this scheme is not applicable to existing standards which don't have these signatures.

The designed and implemented classifier presented in this paper covers the major wireless systems which exist or will potentially operate in UHF band below 800\,MHz, they are DVB-T, 3GPP LTE, IEEE 802.22, ECMA-392 and WM signal. We propose the dimension cancelation\,(DIC) scheme which can completely mitigate the noise uncertainty\,(NU) problem\cite{noise_uc:shellhammer} in receiver. Other harmful issues in sensing receiver such as unflat noise floor and spurs are also taken into account and mitigated. 
The proposed classifier is well validated with both computer simulation and experimental measurement using a testbed.

\section{The Signal Classification Scheme}
\label{sec:classi}

\begin{figure}[tbp]
  \centering
  \includegraphics[width=8.6cm]{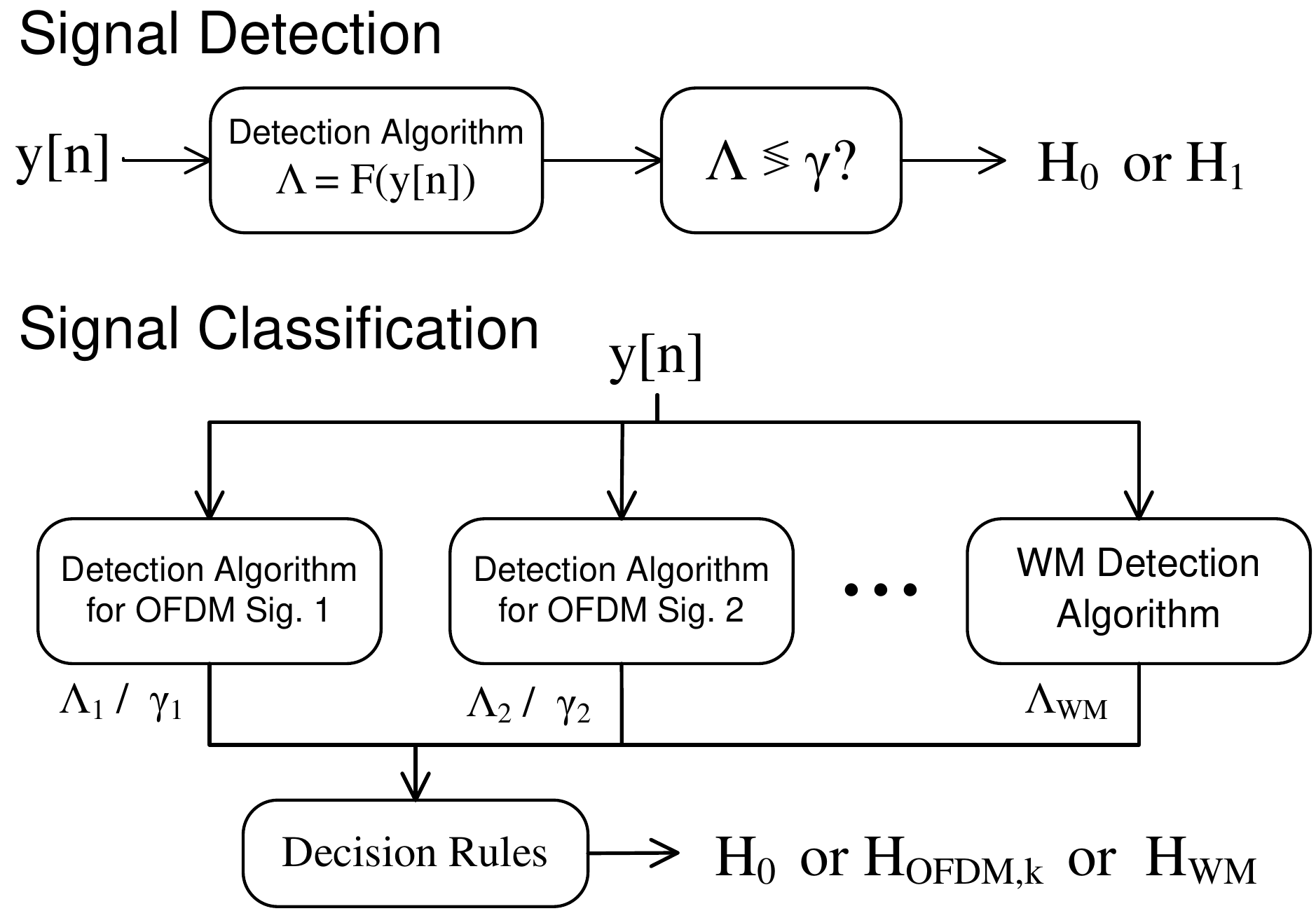}
  \caption{Signal classification based on parallel detection algorithms}
  \label{fig:det_classi_paper}
\end{figure}

The proposed classification scheme is based on combining a group of feature based signal detection algorithms which can test the received signal in parallel\,(Fig.\ref{fig:det_classi_paper}). Unique features are utilized by different detection algorithms, which makes the unmatched signal appears to be like noise. In this way, classification is achieved by matching the detection algorithms to received signal with properly designed decision rules.

\subsection{Targeted Signals \& According Detection Algorithms}

The targeted signals of the classifier are presented in Table \ref{tbl:classi_fea_algo} with their according detection algorithms and utilized features listed. It is well known that the goal of signal detection in spectrum sensing is to identify the following hypotheses:
\begin{equation}
\label{eq:sens_hypo_bas}
\small
\begin{split}
\mathcal{H}_0: y[n] &= w[n] \\
\mathcal{H}_1: y[n] &= x[n] + w[n],
\end{split}
\end{equation}
where $x[n]$ is the signal with channel's effect and $w[n]$ is noise. For detecting the existence of the interested signal $x[n]$, normally, a detection metric $\Lambda$ should be defined, which is the result of certain processing on the received signal $y[n]$. The detection can be achieved by comparing the detection metric $\Lambda$ with a threshold $\gamma$ which is predefined according to a desired probability of false alarm\,(PFA):
\begin{equation}
\label{eq:det_comp}
\Lambda = \mathcal{F}\big{\{} y[n] \big{\}}
\begin{array}{c}
\mathcal{H}_1   \\
\gtreqless      \\
\mathcal{H}_0
\end{array}
\gamma.  
\end{equation}

For detecting DVB-T signal, we adopt a robust detection algorithm called TDSC-MRC 
which was proposed in \cite{dvbt:chen2009} and further studied by us in \cite{dvbt:tdsc:hanwen2010} using real-world TV signal and with practical issues considered. It utilizes the autocorrelations based on the pilot subcarriers' period of four OFDM symbols:
\begin{equation}
\label{eq:fsa_corr}
\small
R^L_k = \frac{1}{\sqrt{L}}\sum_{n=0}^{(N_p-k)L-1} y[n]y^*[n+kL],
\end{equation}
in which $y[n]$ is the received signal, $L$ is the length of one period and $N_p$ is the number of periods within the observation time. Then the detection metric is composed by further correlating the adjacent autocorrelation values $R^L_k$:
\begin{equation}
\label{eq:fsa_2c}
\small
\Lambda_{TDSC-MRC} = \bigg{|}\sum_{k=1}^{J} R^L_k {R^L_{k+1}}^*\bigg{|} \quad J = 1, 2, ..., N_p-2.
\end{equation}
By analyzing the probability distribution of $\Lambda_{TDSC-MRC}$ at $\mathcal{H}_0$ with central limit theorem\,(CLT), the threshold according to the PFA $P_{FA}$ is derived as
\begin{equation}
\label{eq:thr:tdsc_mrc}
\small
\gamma_{TDSC-MRC} = \sigma_{w}^4\sqrt{-\sum_{k=1}^{J} (N_p-k)(N_p-k-1) \ln P_{FA}},
\end{equation}
in which $\sigma_w^2$ is the power of noise $w[n]$.

\begin{table}[tbp]
\caption{Signal Features and Detection Algorithms Used in Classification}
\tiny
\centering
\begin{tabular}{l|l|l}
  \hline
  \hline  \textbf{Standards/Types}      & \textbf{Algorithms}   & \textbf{Utilized Features}\\
  \hline  DVB-T                         & TDSC-MRC\cite{dvbt:chen2009,dvbt:tdsc:hanwen2010}  & periodical pilot tone structure in time domain   \\
  \hline  LTE (5\,MHz mode)             & CP-SW (pre-align.)\cite{sens:dvbt:huawei}    & CP and DFT length, slot length    \\
  \hline  IEEE 802.22 (8\,MHz mode)     & CP-SUM\cite{sens:ofdm_cp_sum:chaudhari2009}  & DFT length\,(indifferent to CP length) \\
  \hline  ECMA-392 (8\,MHz mode)        & CP-SUM\cite{sens:ofdm_cp_sum:chaudhari2009}  & DFT length\,(indifferent to CP length) \\
  \hline  WM                            & PAR and DS of PSD  & high PAR \& low DS in PSD due to narrowband \\
  \hline
  \hline
\end{tabular}
\label{tbl:classi_fea_algo}
\end{table}

For LTE signal, we adopt the algorithm utilizing CP based autocorrelation convoluted with a sliding windowed\,(SW)\cite{sens:dvbt:huawei}, which is named CP-SW detection here. First, the autocorrelation with time lag of OFDM's DFT size is performed:
\begin{equation}
\label{eq:cp_corr}
\small
r[n] = y[n]y^*[n+N_{DFT}]. 
\end{equation}
The autocorrelation values $r[n]$ are further aligned according to the OFDM symbol length
\begin{equation}
\label{eq:cp_sec}
\small
R[n] = \sum_{l=0}^{\big{\lfloor}\frac{N-N_{DFT}+1}{N_{DFT}+N_{CP}}\big{\rfloor}-1}r[n+l(N_{DFT}+N_{CP})]\quad n = 0,\ldots,N_{DFT}+N_{CP}-1.
\end{equation}
in which $N$ is the total number of samples in the observation time. The detection metric of CP-SW is then formulated as
\begin{equation}
\label{eq:cp_sw}
\small
\Lambda_{CP-SW}=\max_{i} \Bigg|\sum_{n=i}^{i+N_{CP}-1}\tilde{R}[n]\Bigg| \quad i = 0, 1, ..., N_{DFT}+N_{CP}-1,
\end{equation}
where $\tilde{R}[n]$ is the cyclic extension of $R[n]$ with CP length. If the signal has different CP lengths, such as the normal CP mode of LTE in which the first symbol of a slot has longer CP length than the other symbols, we propose to use pre-alignment on $r[n]$ according to the periodical frame length:
\begin{equation}
\label{eq:cp_sw_prealign}
\small
r'[n] = \sum_k r[n+kL_{frm}],
\end{equation}
in which $L_{frm}$ is the length of one periodical frame, for LTE it is the length of one slot. The probability distribution of $\Lambda_{CP-SW}$ in $\mathcal{H}_0$ is difficult to obtain, hence we cannot derive the closed-form expression of the threshold $\gamma_{CP-SW}$. However, it can still be estimated through empirical method with large amount of detection tests in hypothesis $\mathcal{H}_0$.

For ECMA-392 which use CSMA/CA MAC,
the signal behaves as random bursts and has no periodical structure within the observation time. However, the detection metric can still be composed by the summation of $r[n]$\cite{sens:ofdm_cp_sum:chaudhari2009}:
\begin{equation}
\label{eq:cp_sum}
\small
\Lambda_{CP-SUM} = \frac{1}{\sqrt{N-N_{DFT}+1}}\bigg|\sum_{n=0}^{N-N_{DFT}-1}r[n]\bigg|.
\end{equation}
By analyzing the probability distribution of $\Lambda_{CP-SUM}$ using CLT, we derive the threshold
\begin{equation}
\label{eq:thr:cp_sum}
\small
\gamma_{CP-SUM} = \sigma_{w}^2 \sqrt{-\ln P_{FA}}.
\end{equation}

For WM signal, we make use of its narrowband\,($<$ 200\,KHz) feature exhibiting sparse distribution and high peak-to-average ratio\,(PAR) in estimated PSD of the signal in one TV channel. The PSD is estimated using the Welch's method:
\begin{equation}
\label{eq:psd_welch_avg}
\small
\hat{Y}[m] = \frac{\sum_{i=0}^{K-1}\Big{|}\sum_{n=0}^{M-1}y[iD+n]v[n]e^{-2\pi j \frac{nm}{M}}\Big{|}^2}{K\sum_{n=0}^{M-1}v^2[n]}
\quad m = 0,1,...,M-1,
\end{equation}
where $M$ is the DFT size, $D$ is the shifting step of DFT segment and the smoothing window $v[n]$ of rectangular window is choose for achieving good frequency resolution. The PAR of the estimated PSD is calculated as

\begin{equation}
\label{eq:det_wm_par}
\small
\Phi = \max(\hat{Y}[m]) \,\big{/}\, \overline{\hat{\mathbf{Y}}},
\end{equation}
in which $\overline{\hat{\mathbf{Y}}}$ denotes the mean value of the estimated PSD $\hat{Y}[m]$. Since the targeted OFDM signals may also exhibit high PAR in their PSD if they experience frequency-selective fading, another property called the degree of sparsity\,(DS) is also used together with PAR to classify the WM signal:
\begin{equation}
\label{eq:det_wm_ds}
\small
\Psi = \sum_{\hat{Y}[m] \geq (1+\rho)\overline{\hat{\mathbf{Y}}}}m \,\Big{/}\, M.
\end{equation}
The detection metric for WM signal becomes the logic result that if $\Phi$ and $\Psi$ are both larger than their respective thresholds:
\begin{equation}
\label{eq:det_wm_all}
\small
\Lambda_{WM} = (\Phi \geq \phi \quad AND \quad \Psi \leq \psi),
\end{equation}
which is either $0$ or $1$.

\subsection{Decision Rules}
After the received signal are processed by the detection algorithms resulting detection metrics $\Lambda_i$ for OFDM class $i$ and $\Lambda_{WM}$ for WM, the rules in (\ref{eq:classi_dec_rule}) are used to decide which signal is occupying the channel\,($\hat{\mathcal{H}}_{OFDM,k}$ or $\hat{\mathcal{H}}_{WM}$) or the channel is vacant\,($\hat{\mathcal{H}}_{0}$). The WM signal is classified as long as the PAR and DS satisfies (\ref{eq:det_wm_all}). Since the different OFDM detection algorithms has different threshold, we take $\Lambda_i/\gamma_i$ as unified metrics with common threshold of 1 for finding the matched algorithm. Because each algorithms is specific to certain standards and modes, one signal can be classified as long as its corresponding algorithm is matched.

\begin{equation}
\label{eq:classi_dec_rule}
\small
\begin{split}
\hat{\mathcal{H}}_{OFDM,k}:  &k = arg \max_i(\Lambda_i/\gamma_i), \Lambda_{WM}=0 \; and \; \max_i(\Lambda_i/\gamma_i) \geq 1 \\
\hat{\mathcal{H}}_{WM}:  &\Lambda_{WM}  =  1     \\
\hat{\mathcal{H}}_0:  &\max_i(\Lambda_i/\gamma_i) < 1 \; and \; \Lambda_{WM} = 0
\end{split}
\end{equation}

\section{Mitigation of Receiver Imperfections}

\begin{figure}[tbp]
  \centering
  \includegraphics[width=8.6cm]{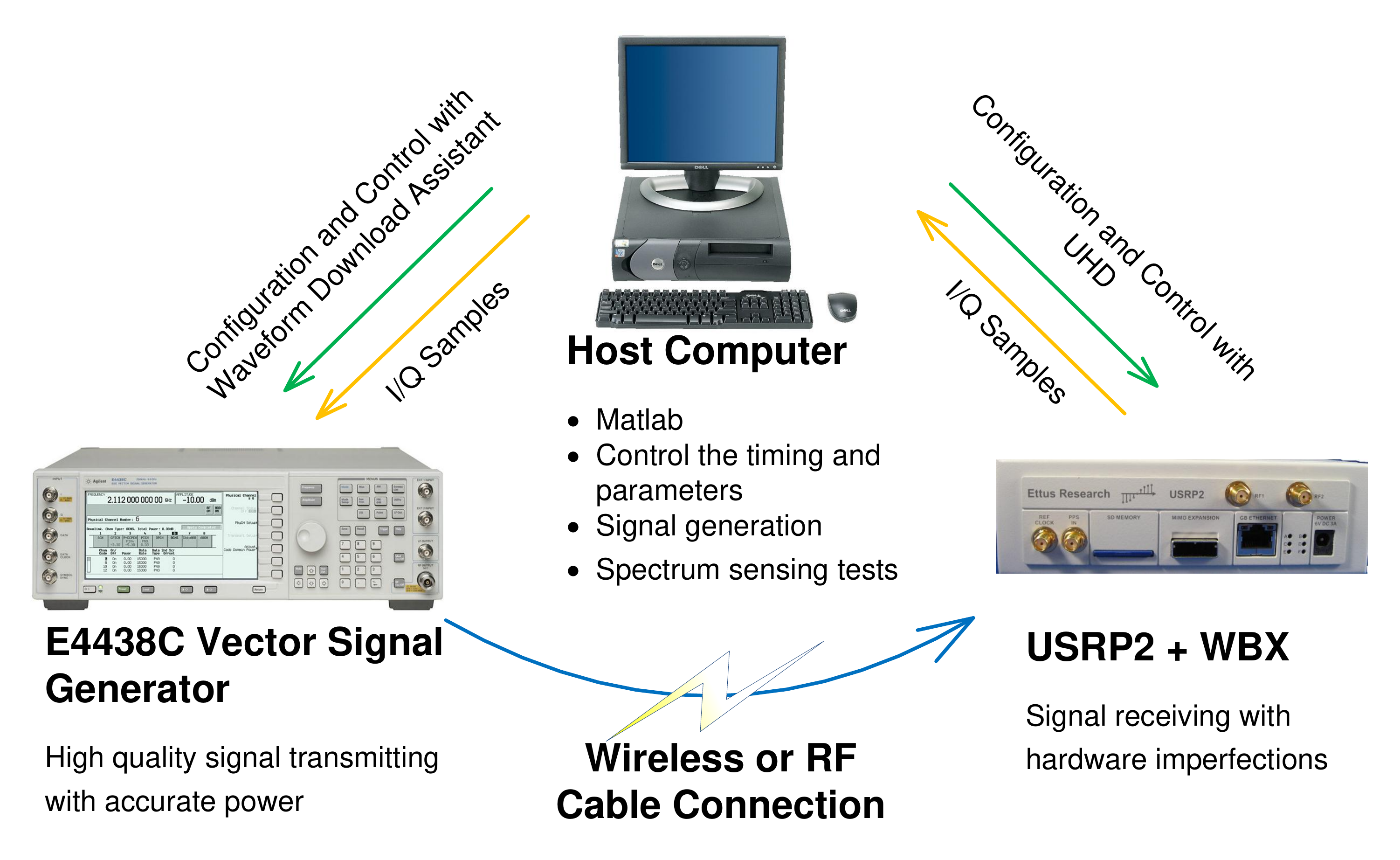}
  \caption{The structure of the spectrum sensing testbed}
  \label{fig:testbed_conf_paper}
\end{figure}

\subsection{Spurs and Unflat Noise Floor}
The narrowband spurs in practical receiver can be mistakenly regarded as WM signal by the classifier. Since the spurs' frequencies are normally fixed, the simple solution is to directly exclude the spurs' frequency components in the estimated PSD $\hat{Y}[m]$ when calculating PAR and DS values in (\ref{eq:det_wm_par}) and (\ref{eq:det_wm_ds}). In the TDSC-MRC, CP-SW and CP-SUM detections which are based on time-domain autocorrelation, the spurs generate strong components in the autocorrelation which can degrade the detection and classification performances. The solution is to use band-stop filters to eliminate the spurs.

The unflat noise floor can lead to nonuniform sensing performances for WM signals at different center frequencies. The simple solution is to pre-estimate the noise PSD $\hat{W}[m]$ at $\mathcal{H}_0$ and use it to equalize the estimated signal's PSD: $\hat{Y}_{eq.}[m] = \hat{Y}[m] / \hat{W}[m]$ during the sensing tests. Then the $\hat{Y}_{eq.}[m]$ is used for further calculation of PAR and DS values. It is found that the autocorrelation based detection algorithms are not sensitive to the unflatness of noise floor.

\subsection{DIC for Mitigating Noise Uncertainty Problem}
It can be noticed in (\ref{eq:thr:tdsc_mrc}) and (\ref{eq:thr:cp_sum}) that the thresholds for TDSC-MRC and CP-SUM aglorithms have the multiplication term $\sigma_{w}^4$ or $\sigma_{w}^2$. This means that the knowledge of noise power $\sigma_{w}^2$ is required in these algorithms. Practically, since $\sigma_{w}^2$ cannot be exactly known due to the NU problem, the thresholds therefore become inaccurate, which leads to unpredicted PFA and probability of mis-detection\,(PMD).

We found that using the new detection metric
\begin{equation}
\label{eq:metric_dic}
\small
\Lambda' = \Lambda\,\big{/}\,\big{(}\hat{\sigma}_y^2\big{)}^\alpha
\end{equation}
instead of the original metric $\Lambda$, the knowledge of noise power $\sigma_{w}^2$ is not required. The $\hat{\sigma}_y^2 = \frac{1}{N}\sum_{n=0}^{N-1}|y[n]|^2$ is the estimated power of the received signal.
The $\alpha$ is the factor which makes the denominator has the same dimension as the nominator in (\ref{eq:metric_dic}). For example, assuming the dimension of received samples $y[n]$ is in volt\,(V), the detection metric $\Lambda_{TDSC-MRC}$ in (\ref{eq:fsa_2c}) is $V^4$, which requires $\alpha=2$. The metrics $\Lambda_{CP-SUM}$ in (\ref{eq:cp_sum}) and $\Lambda_{CP-SW}$ in (\ref{eq:cp_sw}) have dimension of $V^2$, hence, they require $\alpha=1$. This method is called dimension cancellation\,(DIC) in this paper. Since the PAR metric for WM signal is dimensionless, it is inherently unaffected by NU and the DIC is not needed.

\section{Simulation \& Experimental Results}

The configuration of the spectrum sensing testbed is presented in Fig.\ref{fig:testbed_conf_paper}. As listed in Table \ref{tbl:classi_eval_sig_mode}, five classes of signals with various modes are targeted in the tests. These signals are generated in Matlab according to their standard specifications emphasizing the features utilized by the classifier. They are also transmitted by E4438C in the sensing experiments. 

\begin{table}[tbp]
\caption{The Signals and Modes in Classification Test}
\tiny
\centering
\begin{tabular}{l|l}
  \hline
  \hline  \textbf{Standards/Types}      & \textbf{Modes}                                 \\
  \hline  DVB-T                         & 8 modes: all the CP and DFT lengths                \\
  \hline  TD-LTE DL (5\,MHz mode)       & 2 modes: TD-LTE (UL/DL conf. 2), long CP and normal CP    \\
  \hline  802.22 DL (8\,MHz mode)       & 1 mode: 8\,MHz, 1/16 CP                \\
  \hline  ECMA-392 (8\,MHz mode)        & 1 mode: 8\,MHz, 1/16 CP, duty cycle: 0.5                \\
  \hline  WM                            & 1 mode: FM, loud speaker           \\
  \hline
  \hline
\end{tabular}
\label{tbl:classi_eval_sig_mode}
\end{table}

Three methods are employed for evaluating performances:
\begin{itemize}
  \item \emph{Method 1}(M1): Use the signal and white noise generated in Matlab for simulation tests.
  \item \emph{Method 2}(M2): The signal is transmitted by E4438C and captured by USRP2+WBX with high SNR, the noise with spurs is captured by USRP2+WBX. Then they are combined according to the desired SNR in simulations.
  \item \emph{Method 3}(M3): Similar to M2, however, the received signal with local noise and spurs is directly used for sensing tests which features the real spectrum sensing scenario. The SNR is controlled using the ability of E4438C which can scale its transmitting power accurately.
\end{itemize}

\begin{figure}[tbp]
  \centering
  \includegraphics[width=8.6cm]{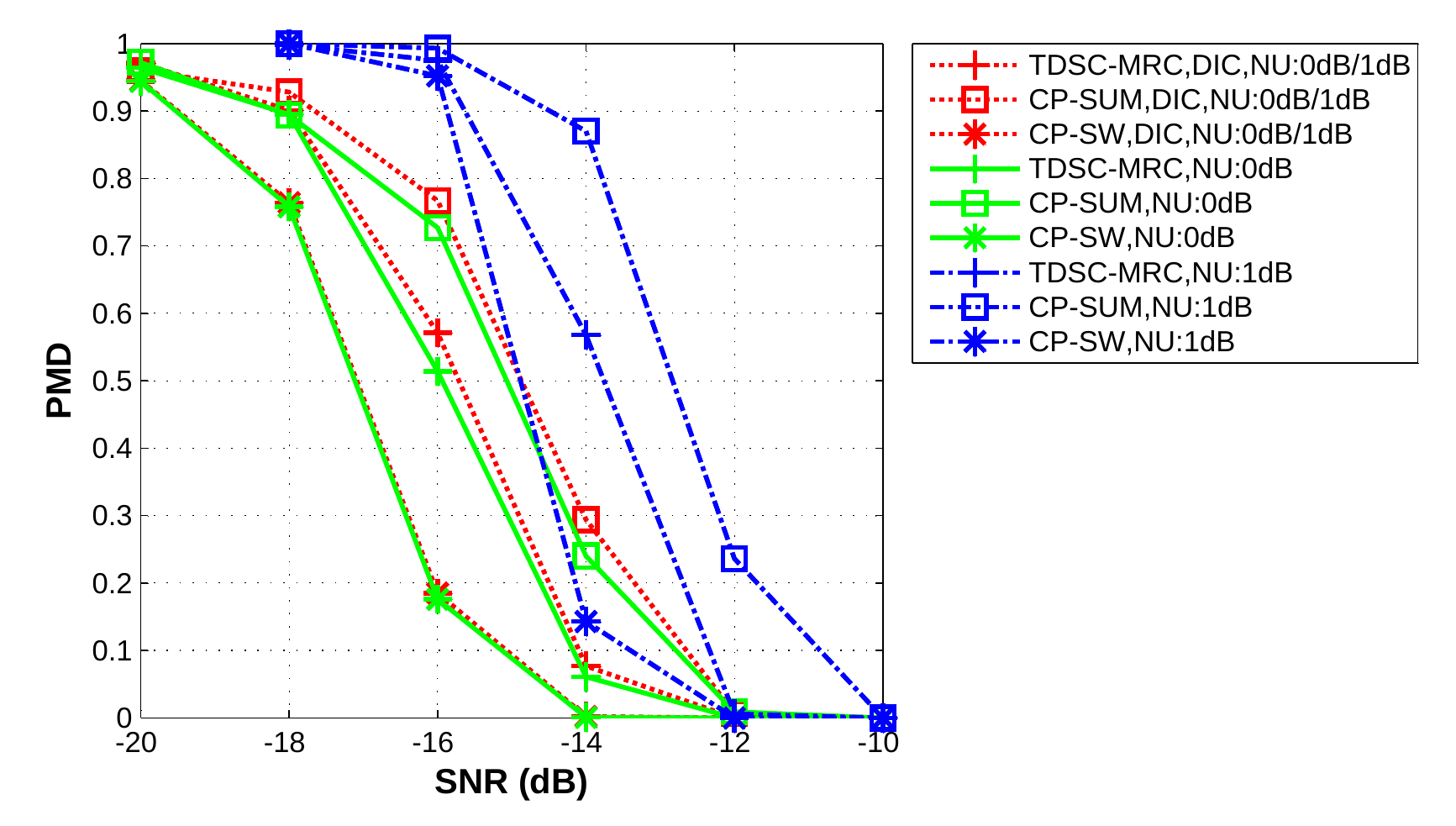}
  \caption{Using DIC to mitigate NU for TDSC-MRC and CP based detections on DVB-T signal\,(2K, 1/4 CP mode), observation time: 10\,ms, PFA: 0.01}
  \label{fig:plot_dvbt_fsa_cp_nu_paper}
\end{figure}

We first validate the effectiveness of applying DIC to TDSC-MRC, CP-SW and CP-SUM detections for mitigating NU problem with simulation. The NU is modeled using the ``robust statistics'' method\,\cite{noise_uc:shellhammer}, in which the upper limit of noise power is used to calculate the PFA while the lower limit is used to calculate the PMD. The simulation performances are presented in Fig.\ref{fig:plot_dvbt_fsa_cp_nu_paper} which shows that the PFA performances of the three algorithms are notably degraded by NU of 1\,dB. However, when DIC is applied, the detection performances become invulnerable to NU at all and are very close to the performances in the ideal case without DIC and NU.

In the signal classification experiment, the sampling rate of the USRP2+WBX is set to 12.5\,MS/s which can well accommodate one TV channel of 8\,MHz.
The signals are transmitted at center frequency of 560\,MHz in which we've acquired transmission license from the local regulator. In the classification algorithm, we use observation time of 20\,ms resulting $12.5MS/s \times 20ms = 0.25M$ samples being processed.

For classifying WM signal, DFT length $M=256$ and shifting step $D=M/2$ is chosen for estimating the PSD. We set the PAR threshold $\Phi=2$, the DS threshold $\Psi=0.2$ and $\rho=0.2$ in (\ref{eq:det_wm_par}) and (\ref{eq:det_wm_ds}) based on empirical tests. For classifying the OFDM signals, the thresholds are obtained via both analytical calculation in (\ref{eq:thr:tdsc_mrc}) and (\ref{eq:thr:cp_sum}) as well as empirical tests with large amount of noise samples according to $PFA$ of 0.01. The fixed sampling rate of 12.5\,MS/s can introduce some out-of-band noise and signal which interfere the classification. Hence, before being processed by the detection algorithms for different OFDM standards, the received signal is filtered and resampled according to their standard-specific sampling rate.

The DIC is applied to all the OFDM detection algorithms. Besides, in M2 and M3 using the USRP2+WBX, the spurs at 560\,MHz and 562.5\,MHz are removed by excluding their PSD components in the WM detection algorithm and by using band-stop filter in the OFDM detection algorithms.

\begin{figure}[tbp]
  \centering
  \includegraphics[width=8.6cm]{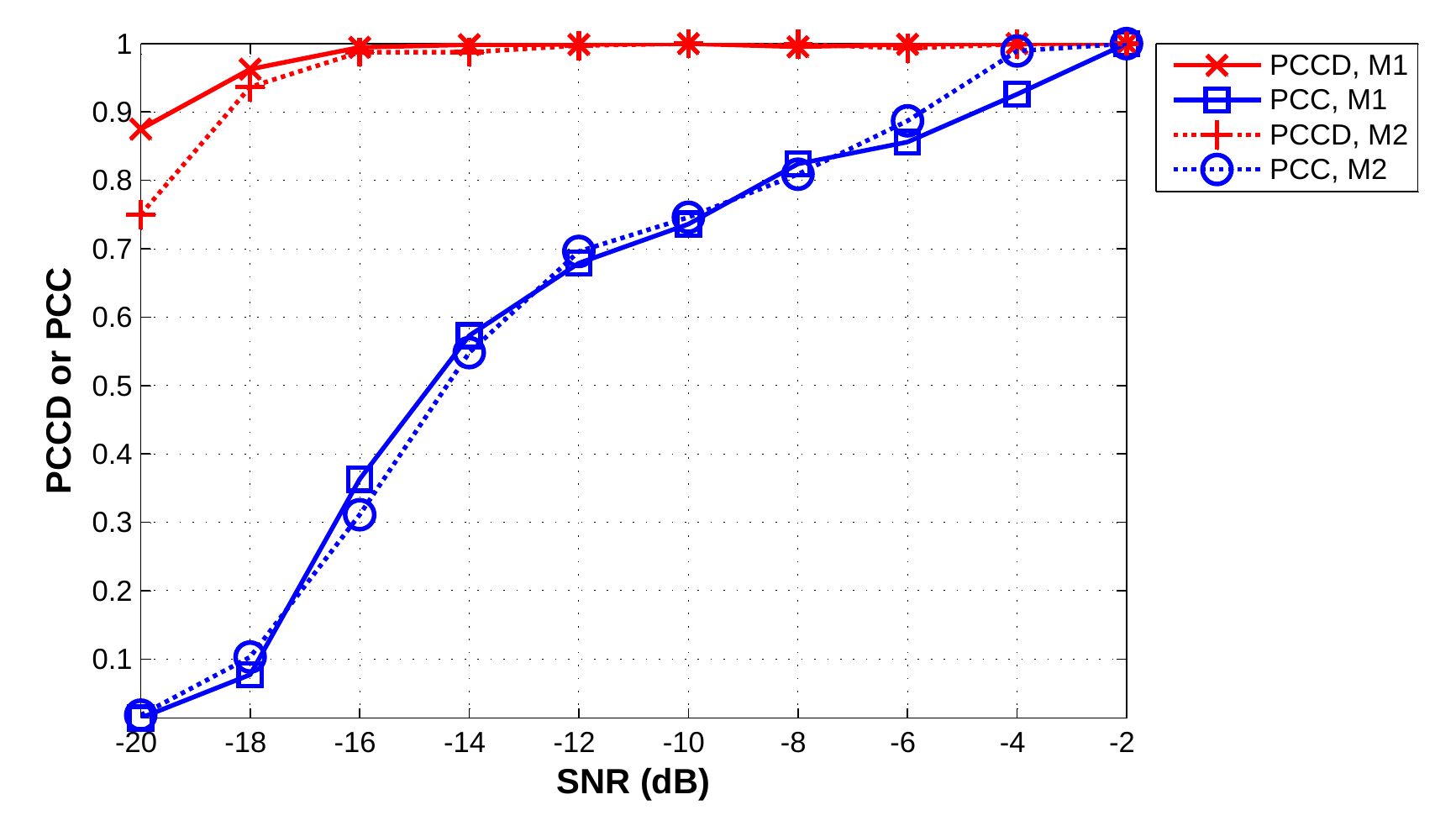}
  \caption{Overall classification performances of all the five signal classes\,(Table \ref{tbl:classi_eval_sig_mode}) using M1, M2, observation time: 20\,ms, PFA: 0.01}
  \label{fig:plot_classi_cap_sim_all}
\end{figure}

In performance evaluations, the five signal classes and their different modes are randomly chosen and processed by the classifier which gives the result of $\hat{\mathcal{H}}_{OFDM,k}$, $\hat{\mathcal{H}}_{WM}$ or $\hat{\mathcal{H}}_0$. The overall classification performance for all the signal classes are presented in Fig.\ref{fig:plot_classi_cap_sim_all}. The probability of correct classification\,(PCC) characterizes how probable the existence of the signal is detected with its type correctly classified. Differently, the probability of correct classification when detected\,(PCCD) characterizes the conditional probability of correct classification when the signal is detected. In Fig.\ref{fig:plot_classi_cap_sim_all}, the PCCD is close or equal to 1 at all the tested SNRs which shows a favorable property of the classifier that as long as the targeted signals are detected, they can be always correctly classified.

\begin{figure}[tbp]
  \centering
  \includegraphics[width=8.6cm]{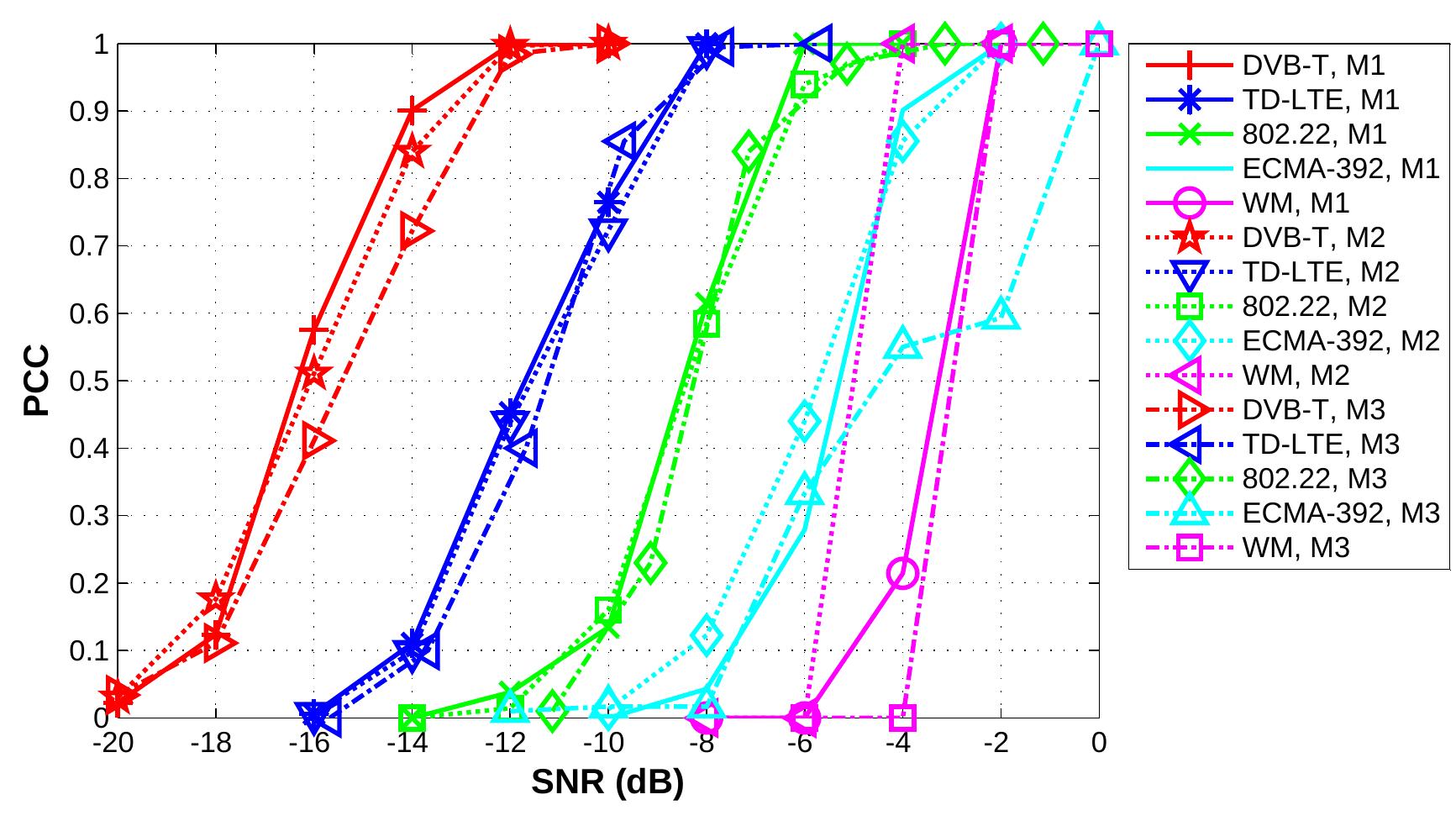}
  \caption{Classification performances of the five individual signal classes\,(Table \ref{tbl:classi_eval_sig_mode}) using M1, M2 and M3, observation time: 20\,ms, PFA: 0.01}
  \label{fig:plot_classi_cap_sim_mea_all}
\end{figure}

Fig.\ref{fig:plot_classi_cap_sim_mea_all} further presents the PCC performances of the five individual signal classes. It shows good agreement between the simulation results using M1/M2 and the experimental results using M3 for all the signal classes, which well validate the effectiveness and feasibility of the proposed classification scheme in real-world spectrum sensing scenario. It also shows that the methods for mitigating spurs and unflat noise floor used in M2 and M3 cause little degradation to classification performance by comparing with the simulation results in M1 with ideal signal and white noise.

\section{Conclusion}

This paper presents a novel signal classification scheme configured for identifying the major existing and emerging wireless systems in UHF band below 800\,MHz. The DIC method is proposed which can completely eliminate the influence of noise uncertainty problem. Apart from simulation, the classification algorithm is implemented on a spectrum sensing testbed considering practical issues of spurs and unflat noise floor. The experimental results agree well with the simulation results, which successfully validates the the effectiveness and feasibility of the proposed signal classification scheme.

\setstretch{0}
\scriptsize
\bibliographystyle{IEEEtran}
\bibliography{IEEEabrv,classi_paper}

\begin{thebibliography}{10}
\providecommand{\url}[1]{#1}
\csname url@samestyle\endcsname
\providecommand{\newblock}{\relax}
\providecommand{\bibinfo}[2]{#2}
\providecommand{\BIBentrySTDinterwordspacing}{\spaceskip=0pt\relax}
\providecommand{\BIBentryALTinterwordstretchfactor}{4}
\providecommand{\BIBentryALTinterwordspacing}{\spaceskip=\fontdimen2\font plus
\BIBentryALTinterwordstretchfactor\fontdimen3\font minus
  \fontdimen4\font\relax}
\providecommand{\BIBforeignlanguage}[2]{{%
\expandafter\ifx\csname l@#1\endcsname\relax
\typeout{** WARNING: IEEEtran.bst: No hyphenation pattern has been}%
\typeout{** loaded for the language `#1'. Using the pattern for}%
\typeout{** the default language instead.}%
\else
\language=\csname l@#1\endcsname
\fi
#2}}
\providecommand{\BIBdecl}{\relax}
\BIBdecl

\bibitem{web:cpmse}
``{C-PMSE Project: www.c-pmse.research-project.de}.''

\bibitem{classi:psd_gi:wang2010}
H.~Wang \emph{et~al.}, ``{Blind standard identification with bandwidth shape
  and GI recognition using USRP platforms and SDR4all tools},'' \emph{Cognitive
  Radio Oriented Wireless Networks \& Communications (CROWNCOM), 2010
  Proceedings of the Fifth International Conference on}, pp. 1--5.

\bibitem{cycl:ofdm_classi_caf:han2008}
N.~Han, G.~Zheng, S.~H. Sohn, and J.~M. Kim, ``{Cyclic Autocorrelation Based
  Blind OFDM Detection and Identification for Cognitive Radio},'' in \emph{2008
  4th International Conference on Wireless Communications, Networking and
  Mobile Computing}.\hskip 1em plus 0.5em minus 0.4em\relax IEEE, Oct. 2008,
  pp. 1--5.

\bibitem{cyc:chn_alloc:oener}
M.~Oner and F.~Jondral, ``{On the extraction of the channel allocation
  information in spectrum pooling systems},'' \emph{IEEE Journal on Selected
  Areas in Communications}, vol.~25, no.~3, pp. 558--565, Apr. 2007.

\bibitem{cycl:classi_ofdm_scld:punchihewa}
A.~Punchihewa \emph{et~al.}, ``{On the Cyclostationarity of OFDM and Single
  Carrier Linearly Digitally Modulated Signals in Time Dispersive Channels:
  Theoretical Developments and Application},'' \emph{IEEE Transactions on
  Wireless Communications}, vol.~9, no.~8, pp. 2588--2599, Aug. 2010.

\bibitem{cycl:classi_lte_wimax:al-habashna}
A.~Al-Habashna \emph{et~al.}, ``{Second-Order Cyclostationarity of Mobile WiMAX
  and LTE OFDM Signals and Application to Spectrum Awareness in Cognitive Radio
  Systems},'' \emph{IEEE Journal of Selected Topics in Signal Processing},
  vol.~6, no.~1, pp. 26--42, Feb. 2012.

\bibitem{cyc:sutton_thesis:2008}
P.~D. Sutton, ``{Rendezvous and Coordination in OFDM-based Dynamic Spectrum
  Access Networks},'' Ph.D. dissertation, University of Dublin, Trinity
  College, 2008.

\bibitem{noise_uc:shellhammer}
S.~Shellhammer and R.~Tandra, ``{IEEE 802.22-06/0134r0: Performance of the
  Power Detector with Noise Uncertainty},'' Tech. Rep., 2006.

\bibitem{dvbt:chen2009}
H.-S. Chen, W.~Gao, and D.~Daut, ``{Spectrum Sensing for OFDM Systems Employing
  Pilot Tones},'' \emph{IEEE Transactions on Wireless Communications}, vol.~8,
  no.~12, pp. 5862--5870, 2009.

\bibitem{dvbt:tdsc:hanwen2010}
H.~Cao, S.~Daoud, A.~Wilzeck, and T.~Kaiser, ``{Practical issues in spectrum
  sensing for multi-carrier system employing pilot tones},'' in \emph{3rd
  International Workshop on Cognitive Radio and Advanced Spectrum Management
  (CogArt 2010)}.\hskip 1em plus 0.5em minus 0.4em\relax Rome, Italy: IEEE,
  Nov. 2010, pp. 1--5.

\bibitem{sens:dvbt:huawei}
``{IEEE 802.22-06/0127r0: Sensing Scheme for DVB-T},'' Huawei Technologies,
  UESTC, Tech. Rep., 2006.

\bibitem{sens:ofdm_cp_sum:chaudhari2009}
S.~Chaudhari \emph{et~al.}, ``{Autocorrelation-Based Decentralized Sequential
  Detection of OFDM Signals in Cognitive Radios},'' \emph{IEEE Transactions on
  Signal Processing}, vol.~57, no.~7, pp. 2690--2700, Jul. 2009.

\end{thebibliography}
\end{document}